\documentclass[review]{elsarticle}

\usepackage{siunitx}

\journal{NIM}

\begin{document}

\begin{frontmatter}

\title{Upgrade of Hardware Controls for the STAR Experiment at RHIC}

\author[instCreighton]{Jaroslav Adam\fnref{noteAuth}\corref{corrAuth}}
\cortext[corrAuth]{Corresponding author}
\fntext[noteAuth]{Now at Brookhaven National Laboratory, Upton, United States}
\ead{jaroslavadam299@gmail.com}

\author[instCreighton]{Michael G. Cherney}
\author[instCreighton]{Joey D'Alesio}
\author[instCreighton]{Emma Dufresne}
\author[instCTU]{Luk\'{a}\v{s} Holub}
\author[instCreighton]{Janet E. Seger}
\author[instCreighton]{David Tlust\'{y}}

\address[instCreighton]{Creighton University, Omaha, United States}
\address[instCTU]{Czech Technical University in Prague, FNSPE, Prague, Czech Republic}

\begin{abstract}
The STAR experiment has been delivering significant physics results for more than 20 years.
Stable operation of the experiment was achieved by using a robust controls system based on the Experimental Physics 
and Industrial Control System (EPICS). Now an object-oriented approach with Python libraries, adapted
for EPICS software, is going to replace the procedural-based EPICS C libraries previously used at STAR. Advantages
of the new approach include stability of operation, code reduction and straightforward project documentation.
The first two sections of this paper introduce the STAR experiment, give an overview of the EPICS architecture,
and present the use of Python for controls software. Specific examples, as well as upgrades of user interfaces,
are outlined in the following sections.

\end{abstract}

\begin{keyword}
EPICS, Python, PyEpics, PythonSoftIoc
\end{keyword}

\end{frontmatter}

\section{Introduction}
The Solenoidal Tracker at RHIC (STAR) experiment \cite{Radhakrishnan:2019ovu} at the Relativistic Heavy Ion Collider (RHIC)
was designed to study strongly
interacting matter at the highest energy densities. It is equipped for precise tracking and identification of charged particles,
electromagnetic
calorimetry at central angles, measurements of charged particle yields at forward and backward angles, and neutron calorimetry
at very forward and backward directions \cite{Ackermann:2002ad}.

\begin{figure}
\centering
\includegraphics[width=0.9\textwidth]{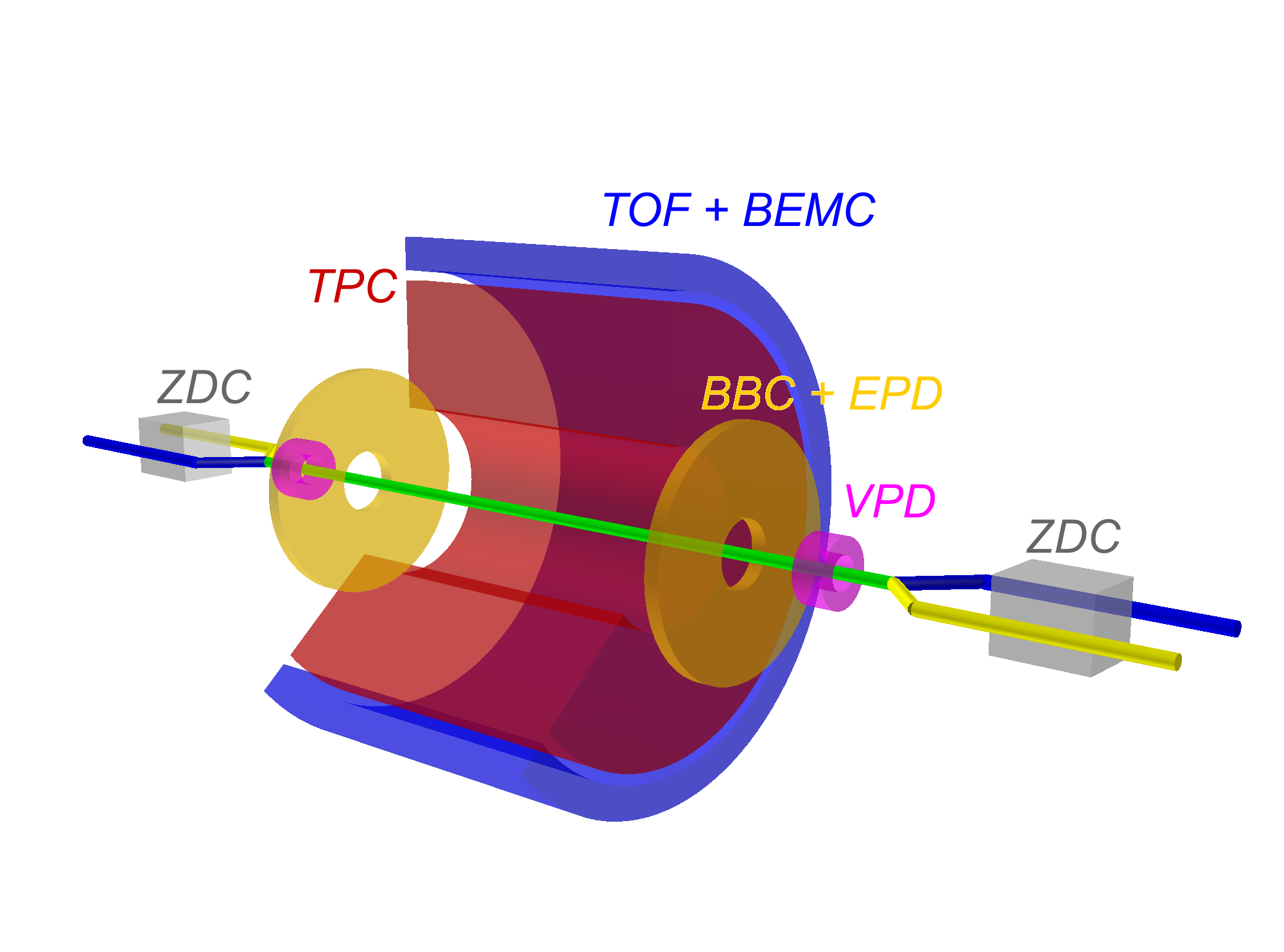}
\caption{STAR experiment at RHIC.}
\label{fig:star}
\end{figure}

Fig.~\ref{fig:star} shows the components of the STAR experiment.
Central tracking, particle identification, and electromagnetic calorimetry are provided by a large gaseous Time Projection Chamber (TPC),
Time-Of-Flight
measurement (TOF) and by the Barrel Electromagnetic Calorimeter (BEMC). Forward charged particles are detected by several scintillator
detectors: the Event Plane Detector (EPD), the Beam-Beam Counter (BBC) and the Vertex Position Detector (VPD). Very forward neutrons
are measured by the Zero Degree Calorimeters (ZDC).

As the experiment is located in a high radiation area, remote control for all the hardware was developed along
with design of the experiment \cite{Reichhold:2003af, Lin:2000vx, Gross:1993jm}.
It is necessary to control and monitor over \si{\num{50000}} parameters of the running experiment. These include voltages and currents
on detection and readout elements, temperatures, power supply parameters, magnetic fields and environmental parameters
in the experimental hall.

\begin{figure}
\centering
\includegraphics[width=0.7\textwidth]{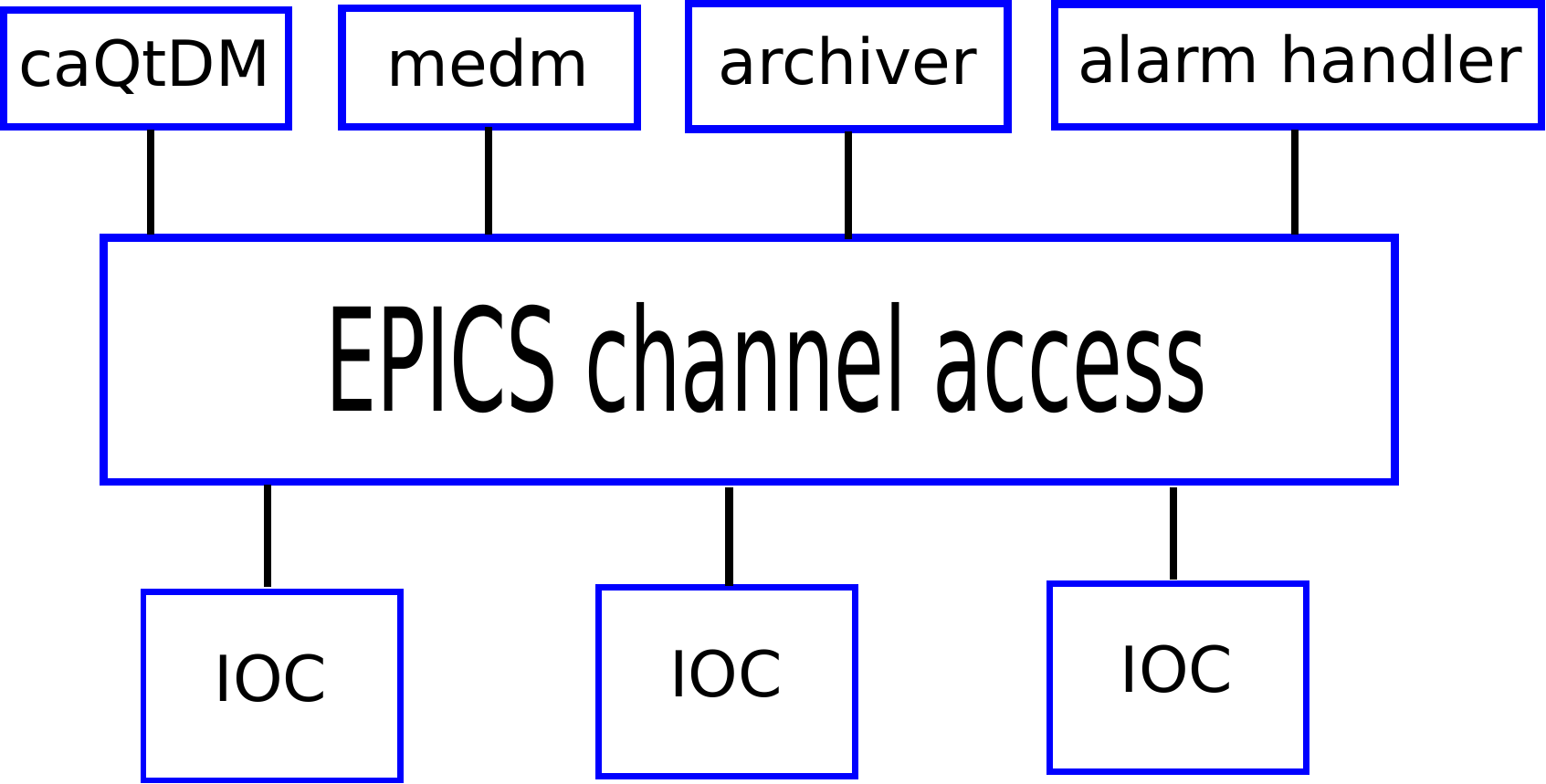}
\caption{Overview of EPICS architecture. The Channel Access protocol provides communication between servers (IOC) and client
applications.}
\label{fig:epics}
\end{figure}

The detector control system is based on the Experimental Physics and Industrial Control System (EPICS) \cite{epics}.
EPICS provides the means to control and monitor the detector parameters via Process Variable (PV) objects that represent
the physical detector parameters in question. The PV objects are managed by the Input-Output Controllers (IOC), which act as
servers in the distributed EPICS system. The PVs can be accessed by a large variety of client applications. Communication
between the IOCs and clients is performed over a local ethernet network using the Channel Access (CA) protocol. All interactions
with physical hardware are provided solely by the IOCs.
Fig.~\ref{fig:epics} gives an overview of the EPICS architecture: several IOCs may be connected via the CA protocol to various
applications such as graphical user interfaces, archiving databases or alarm handlers.

At STAR, the original IOCs were designed using EPICS 3.12, running on Motorola MVME147 and MVME167 single-board processors
with the VxWorks 5.2 operating system. During the experiment's evolution, some of the VxWorks based IOCs were replaced by EPICS 3.14
software IOCs, running on standard PCs under Scientific Linux.
The software IOCs were built upon procedural-based EPICS libraries specific to incorporate communication with hardware
and to manage the process variables.

\section{Python based controls software}
\begin{figure}
\centering
\includegraphics[width=0.7\textwidth]{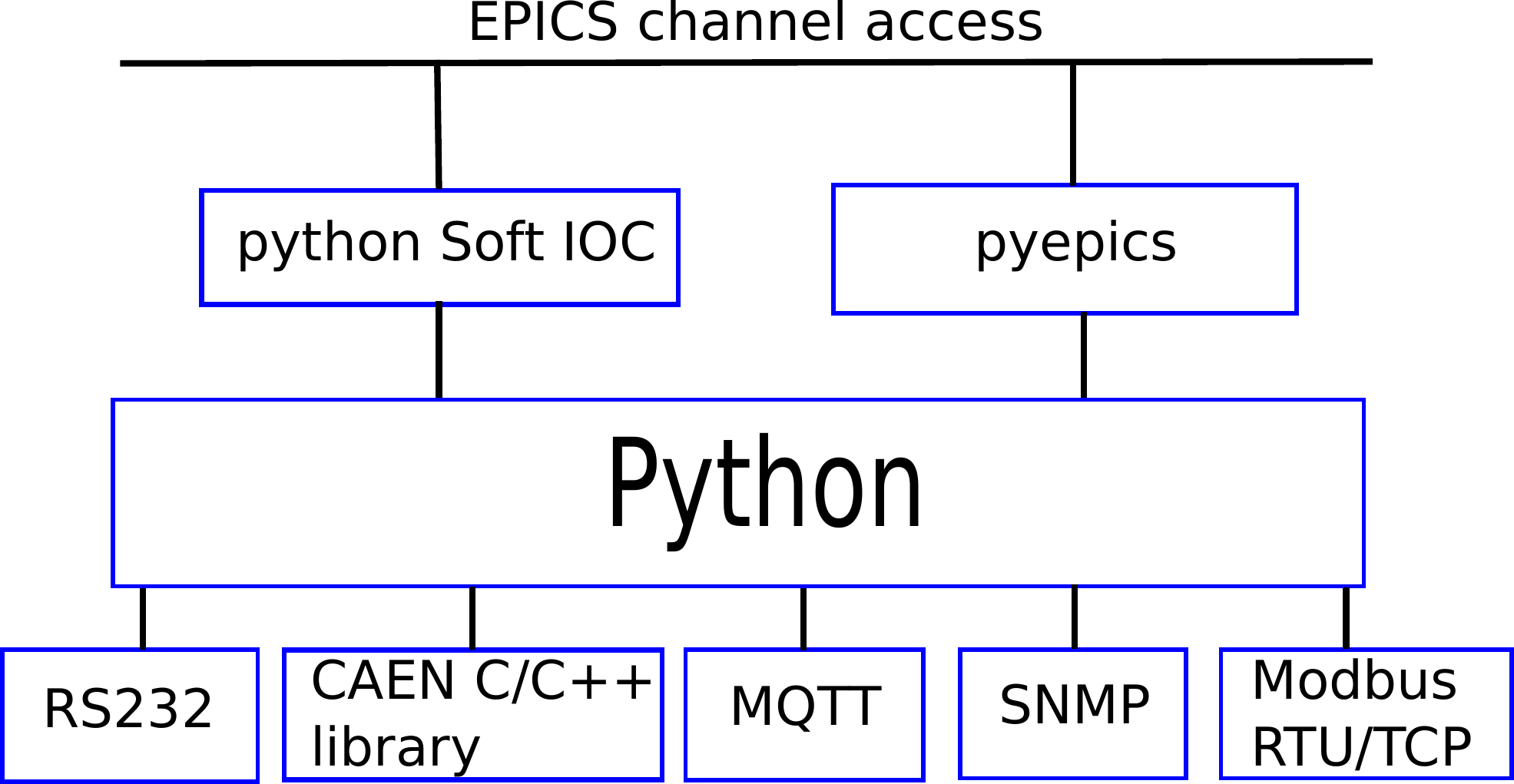}
\caption{Python is software bus between EPICS and hardware. A Python application can create its PV objects via pythonSoftIoc
and connect to PVs of other IOCs via PyEpics. Communication with hardware is achieved by dedicated Python modules.}
\label{fig:pythonIoc}
\end{figure}

EPICS has recently developed tools that are used to create and manage software IOCs entirely in Python \cite{pythonIoc,pyepics}.
The main advantage is a purely object-oriented approach to IOC design which, together with the relative simplicity of the Python
syntax, results in a well-structured and easily maintainable code.
Python supports all standard communication protocols with hardware, which
reduces the need for a specific hardware support. The use of such standard libraries contributes to further code
reduction.
Fig.~\ref{fig:pythonIoc} gives an outline of how the Python environment can interface with both the EPICS control system
and several hardware control libraries.

As the IOCs now consist of a set of object-oriented Python programs only, it is straightforward to keep and document 
them by means of the Git version control system \cite{git:book}.
The STAR experiment has created an official Github repository \cite{star:github} for detector controls, maintaining all the Python-based
IOCs.

\section{IOC implementations with Python}\label{sec:ioc}

\subsection{High voltage for trigger detectors}
The BBC, VPD and ZDC detectors require high voltage for the Photomultiplier tubes (PMTs) that convert optical scintillation responses
to electrical signals. The voltages to individual PMTs are provided by a LeCroy 1440 high voltage system. The communication
is done using the RS232 serial protocol.

The original VxWorks IOC code at STAR had a design flaw, which required the experiment operating crew
to restart that IOC on a daily basis. Therefore, it was decided to build a new Python IOC \cite{star:bbclecroy} using
the pyserial module \cite{pyserial} for RS232 communication.
The new IOC mimics all the capabilities of the old one, namely the ability to load all demand voltages from a set of text
files. In addition, it allows the user to set several voltage profiles for the detector to ensure a straightforward
means of calibration for the detectors.
Robust operation of the new system was achieved during the 2019 data taking period.

\subsection{Gas monitoring}
The central gaseous detectors, the TPC, TOF and Muon Telescope Detector (MTD), require the monitoring of pressure, flow and temperature
of the working gases and their mixing parameters. The measuring hardware transports the data over an isolated network
to one of the detector control PCs where it is written to a set of local ASCII files.

The original EPICS 3.14 IOC used a set of elementary C functions to access these data. However, it was not able to recognize
a connection loss on the side of the gas monitoring hardware.
The problem was overcome by employing a Python based IOC \cite{star:gasread}, which uses the watchdog.observers module \cite{watchdog}
to automatically recognize when new data arrive.
In the case that no new data arrive during a predefined time interval,
the appropriate alarm levels are set to gas-related PVs, thus notifying the operating crews of the problem.

\subsection{EPD integration in the alarm system}\label{sec:epd}
The Event Plane Detector (EPD) was installed to precisely measure the geometrical plane in which each heavy ion collision occurs.
The EPD group chose the MQTT protocol \cite{mqtt-proto} for control. During commissioning, the request was made to include the EPD
in the standard STAR alarm system.

The set of parameters monitored for conditions outside of the normal operating range are voltages, currents and temperatures
on individual detector elements.
In order to make these available to the alarm system, it was necessary to introduce a one-to-one mapping from the MQTT variables
to the EPICS PV objects.
The task was solved by developing a Python IOC \cite{star:epdalarm}, holding the PVs according to the EPD naming convention,
and retrieving the MQTT variables via the paho.mqtt.client module \cite{paho_mqtt}.
Effectively, the IOC acts as a MQTT to Channel Access bridge.

\subsection{Grid leak wall for iTPC upgrade}
The inner sectors of the TPC (iTPC) were upgraded for a more precise readout, allowing the detection of tracks
at previously inaccessible kinematic intervals.

As a result of electron avalanches near the TPC sense wires, large numbers of positive ions are created.
The ions are kept from entering the TPC drift volume by a set of dedicated grid wires.
When introducing the iTPC upgrade, it was necessary to also block the ions from moving between adjacent readout sectors.
This was achieved by placing electrodes at a negative high voltage between the TPC inner and outer sectors. The setup
is commonly referred to as Grid Leak Wall Suppression.
The voltages are provided by the ISEG modules in a Wiener control assembly. Communication to the modules is provided 
by the Wiener MPOD controller via the SNMP protocol.

The task here was to develop the entire IOC. The Python based IOC was created \cite{star:grid_leak}, utilizing the standard SNMP command
line tools:
snmpwalk and snmpset. Access to the SNMP tools from Python is provided via the subprocess standard Python module.
The solution was very simple to implement and proved its robustness during the experiment run in 2019.

\subsection{Air conditioning for TPC and eTOF}\label{sec:tpc_ac}
The gaseous TPC and endcap TOF (eTOF) detectors are held at pressures slightly higher than atmospheric pressure
to avoid air leaking into the detectors. In addition, constant air temperature and humidity and steady air flow
are necessary for the stable operation of both the TPC and the eTOF.

Two new high volume air conditioning units were put in place to provide stable air flow around the TPC and eTOF detectors.
The units communicate the air flow, temperature and humidity values using the ModbusRTU protocol \cite{modbus2004modbus}.
A TCP to Modbus converter is attached
to each unit as a convenient way to provide remote access to the internal RS485 line.

A Python-based IOC is set to read the air parameters and report them via their PVs. Communication to the units
is done by the pyModbusTCP module \cite{modbustcp}.
The IOC is capable of also turning the units on or off by issuing Modbus signals to the internal power relay.

\subsection{eTOF Low Voltage Power Supplies Monitoring}

A group from the CBM collaboration \cite{Senger:2020iki} installed the eTOF system at STAR with already existing 
EPICS PVs for monitoring currents and voltages of the custom made front-end electronics boards.
However, the values of these PVs contain raw ADC values only. Moreover, as currents are measured on these boards 
by Hall probes, the STAR magnetic field affects these values.
The magnetic fields from relays on these boards can also affect the measured current.
Therefore, new PVs have been defined using a Python-based IOC and the numpy package \cite{numpy}. These new PVs
are assigned with currents 
and voltages calculated from the raw ADC values of the pre-defined CBM PVs. Additionally,
the known errors in the current measurements caused by magnetic fields
can be corrected in these new PVs.
The new PVs
are updated based on callback functions on the value
change which guarantees the efficient use of
computing resources.

\subsection{STAR upgrade plan}
An ambitious upgrade is planned for STAR to enhance tracking and calorimetry in the forward region \cite{Yang:2019bjr}.
Tracking in the forward
region will be provided by the small-strip Thin Gap Chamber (sTGC), the Forward Silicon Tracker, and electromagnetic and hadron
calorimetry.
A prototype for the sTGC was installed during the 2019 run period. Readout of the sTGC requires a positive high
voltage for charge multiplication in a gas (as does the TPC). The voltage to the prototype was provided by a set of unused
channels in the CAEN SY1527 used for the TOF. The IOC was written in Python \cite{star:stgc}, employing the ctypes.cdll
Python module to communicate with the CAEN via the vendor specific .so library.

\section{User Interfaces}\label{sec:user}

\subsection{Slow Controls Monitoring Webpage}

\begin{figure}
\centering
\includegraphics[width=0.9\textwidth]{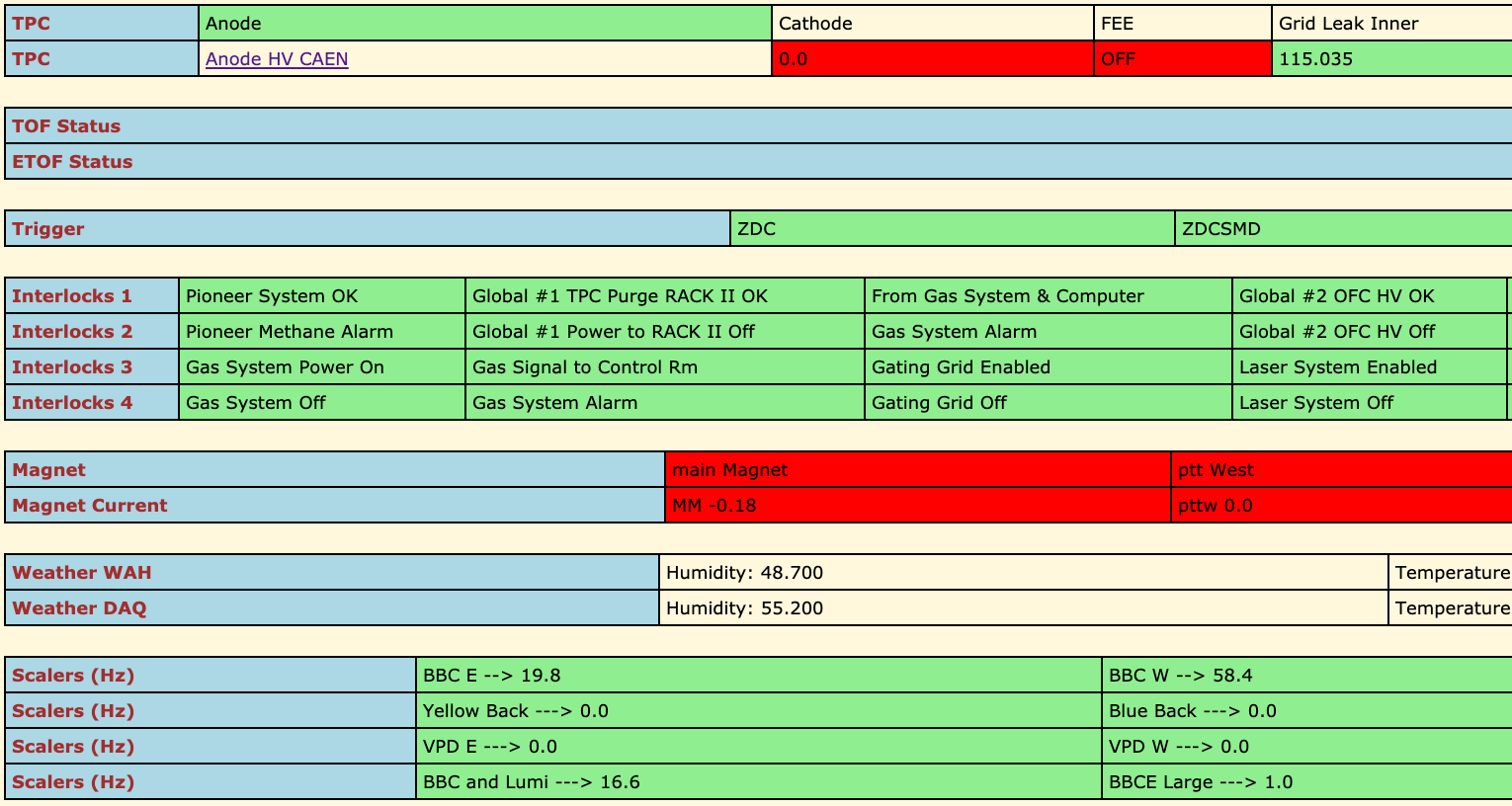}
\caption{A part of the slow controls monitoring web page. Components which are ready to start data taking
are shown in green; systems which are not ready are shown in red.}
\label{fig:webpage}
\end{figure}

The slow controls monitoring webpage, shown in Fig.~\ref{fig:webpage}, allows detector experts and run coordinators outside
the control room to monitor 
information about the detectors during the run. For a quick glance at the major experiment operations, this webpage 
provides information on the run performance and can alert users in case of an operational disruption.
For example, 
information about water and gas alarms, the operating status of the sub-detectors, and environmental conditions can be monitored.
Weather values presented in Fig.~\ref{fig:webpage} are neither red nor green, as their values do not
deter operational validity. They are still important to monitor for potential hazards to the experiment's mechanics.
New code was written using PyEpics to gather the useful information and fill the webpage. With this update, outdated values 
were removed and the use of a legacy unstable software was eliminated. With an object-oriented approach in PyEpics, 
calls to the Channel Access are minimal, allowing for a quick response time.

\subsection{Upgrade of Graphical User Interface}
The Graphical User Interface (GUI) in EPICS is a Channel Access application, aimed at visualizing and/or editing
values of the PVs.

\begin{figure}
\centering
\includegraphics[width=0.7\textwidth]{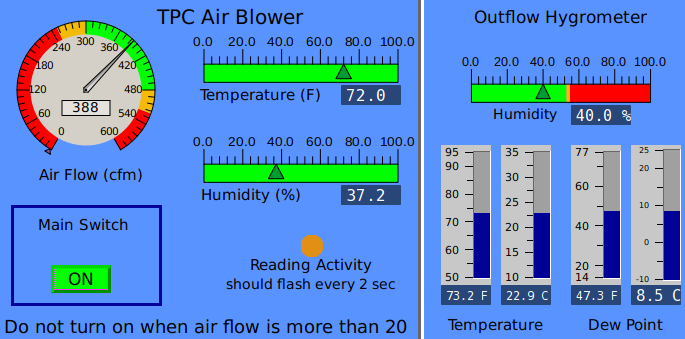}
\caption{An example of a GUI created with caQtDM. The GUI shows the operation of the air conditioning for the TPC and eTOF.}
\label{fig:tpc_ac}
\end{figure}

The Motif Editor and Display Manager (MEDM) has been in use at STAR since the beginning of its operation. Recently, the support
for the motif graphical libraries, upon which the MEDM is built, was discontinued. The situation called for replacement
of the MEDM.
caQtDM \cite{Mezger:ICALEPCS13-TUPPC121} was found to be a suitable replacement for MEDM. It is based upon
the well-supported Qt graphical libraries,
giving the GUIs a fresh modern look. An example of the GUI created with caQtDM,
for the case of the air conditioning for the TPC and eTOF (Section~\ref{sec:tpc_ac}), in shown in Fig.~\ref{fig:tpc_ac}.

An issue might arise when a caQtDM GUI is accessed on a remote machine with X11 forwarding via ssh.
The higher resolution in caQtDM, compared to
simpler shapes in MEDM, requires a larger amount of data to be transmitted.
The GUI was tested in several locations with differing connection speeds, to determine that a basic compression in ssh completely
eliminates any speed issues.

\begin{figure}
\centering
\includegraphics[width=0.8\textwidth]{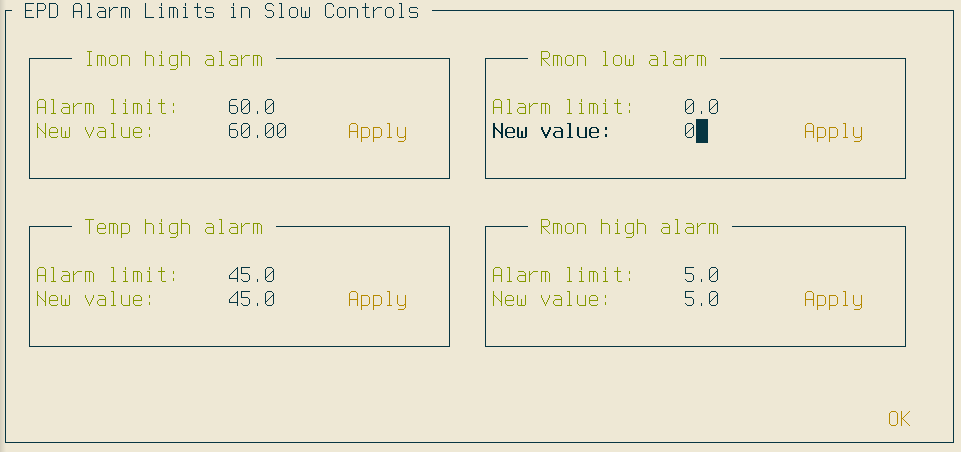}
\caption{An example of an interactive text mode interface. It allows expert users to monitor and/or set alarm levels
related to EPD operation.}
\label{fig:epd_alarm}
\end{figure}

Another approach to controls for experts is based on a text interface rather than a GUI. The npyscreen python module \cite{npyscreen}
makes it possible
to create text mode monitoring and controls objects. When coupled together with PyEpics to manage the Channel
Access, it is straightforward to create an interactive text mode semi-graphical interface, having all the capabilities
of a standard graphical GUI. An example is shown in Fig.~\ref{fig:epd_alarm}. The user can monitor and/or set several
alarm levels related to the operation of the EPD detector (Section~\ref{sec:epd}).

\section{Conclusions}
The STAR experiment has been run for 20 years. Further operation is planned, including an ambitious detector 
upgrade to provide additional tracking and calorimetry in the forward region. This would not be possible without a robust 
control system that is continuously maintained.
Use of an object-oriented Python approach for EPICS software has proven to be a convenient solution to challenges
in the experiment control. The new approach has been successfully deployed to enhance the existing control components,
and is widely accepted at STAR for all intended detector upgrades.

\section*{CRediT authorship contribution statement}
\textbf{J. Adam:} Conception and design of the work, Software development, Writing - original draft.
\textbf{M. Cherney:} Writing - review \& editing, Conception.
\textbf{J. D'Alesio:} Contribution to software, Writing - original draft.
\textbf{E. Dufresne:} Contribution to software, Writing - original draft.
\textbf{L. Holub:} Contribution to software, Writing - original draft.
\textbf{J. Seger:} Supervision, Writing - review \& editing, Conception.
\textbf{D. Tlust\'{y}:} Contribution to software, Writing - original draft.

\section*{Declaration of competing interest}
The authors declare that they have no known competing financial interests or personal relationships that could have
appeared to influence the work reported in this paper.

\section*{Acknowledgment}
The authors thank Jiro Fujita for useful discussions and help.

This work was supported in part by the Office of Nuclear Physics within the US DOE Office of Science.

\section*{References}

\bibliography{references}

\end{document}